# Reversible Logic to Cryptographic Hardware: A New Paradigm

Himanshu Thapliyal and Mark Zwolinski*
Centre for VLSI and Embedded System Technologies
International Institute of Information Technology, Hyderabad, 500032, India
* Electronic System Design Group, Electronics & Computer Science
University of Southampton, UK
(thapliyalhimanshu@yahoo.com, mz@ecs.soton.ac.uk)

*Abstract*— **Differential Power Analysis (DPA) presents a major challenge to mathematically-secure cryptographic protocols. Attackers can break the encryption by measuring the energy consumed in the working digital circuit. To prevent this type of attack, this paper proposes the use of reversible logic for designing the ALU of a cryptosystem. Ideally, reversible circuits dissipate zero energy. Thus, it would be of great significance to apply reversible logic to designing secure cryptosystems. As far as is known, this is the first attempt to apply reversible logic to developing secure cryptosystems. In a prototype of a reversible ALU for a crypto-processor, reversible designs of adders and Montgomery multipliers are presented. The reversible designs of a carry propagate adder, four-to-two and five-to-two carry save adders are presented using a reversible TSG gate. One of the important properties of the TSG gate is that it can work singly as a reversible full adder. In order to design the reversible Montgomery multiplier, novel reversible sequential circuits are also proposed which are integrated with the proposed adders to design a reversible modulo multiplier. It is intended that this paper will provide a starting point for developing cryptosystems secure against DPA attacks.**

## I. INTRODUCTION

Side Channel attacks against cryptographic systems exploit physical characteristics of a device, rather than direct code-breaking methods. One such technique is Differential Power Analysis (DPA), which uses the power consumption of a cryptographic device such as a smartcard [1,2,3]. It is known that the amount of power consumed by the device varies depending on the data and the instructions performed during different parts of an algorithm's execution, thus an attacker directly observes a device's power consumption. By simply examining power consumption traces, it is possible to determine the characteristics of a cryptographic device and the key of the cryptographic algorithm being used. In this work, we propose the use of reversible logic to thwart attacks against cryptographically secure hardware based on DPA. Researchers have shown that for *irreversible* logic computations, each bit of lost information generates $kT\ln 2$ joules of heat energy, where $k$ is Boltzmann's constant and $T$ is the absolute temperature at which the computation is performed [4]. *Reversible* circuits do not lose information, and thus $kT\ln 2$ joules of heat energy will *not* be dissipated [5]. Furthermore, voltage-coded logic signals have an energy of $E_{sig} = \frac{1}{2}CV^2$, and this energy is dissipated whenever the node voltage changes in the irreversible CMOS technology. It is estimated that reversible logic also helps to save energy by using charge recovery logic [7]. Younis has fabricated an 8x8 reversible multiplier array using SCRL gates and measured an energy saving of over 99% over conventional CMOS implementations of the same circuits [8]. Thus, the application of reversible logic to the field of hardware cryptography is proposed here to guard against DPA attacks, as, ideally, no energy will be dissipated in the reversible circuits. Addition and modulo multiplication are the two major power hungry operations in the ALU of a crypto-processor. Thus, this paper proposes a reversible carry propagate adder, four-to-two and five-to-two carry save adders (CSA) using a reversible TSG gate [9,10,11,12]. Furthermore, a reversible Montgomery multiplier [13] using the proposed reversible adders is shown. The major requirement for a Montgomery multiplier is the design of reversible sequential components, thus the authors have also proposed the reversible sequential components like latch, flip flop, register and shift register using the Fredkin gate. The proposed reversible circuits form the primitive components of the ALU of a reversible crypto-processor. As far as we know, this is the first attempt to apply reversible logic to designing secure cryptosystems.

## II. REVERSIBLE TSG GATE

In order to implement the reversible designs of the carry propagate, carry save adders and Montgomery multiplication, a basic reversible TSG gate is discussed along with the term 'garbage output'.

### *A. TSG GATE*

A 4×4 *one through* reversible gate called a TS gate or "TSG" is proposed [9,10,11,12]. The term *one through* means that one of the inputs is directly passed as output. The reversible TSG gate is shown in Fig. 1. It can be verified that the input pattern to generate a particular output pattern can be uniquely determined. The TSG gate is able to implement all Boolean functions.

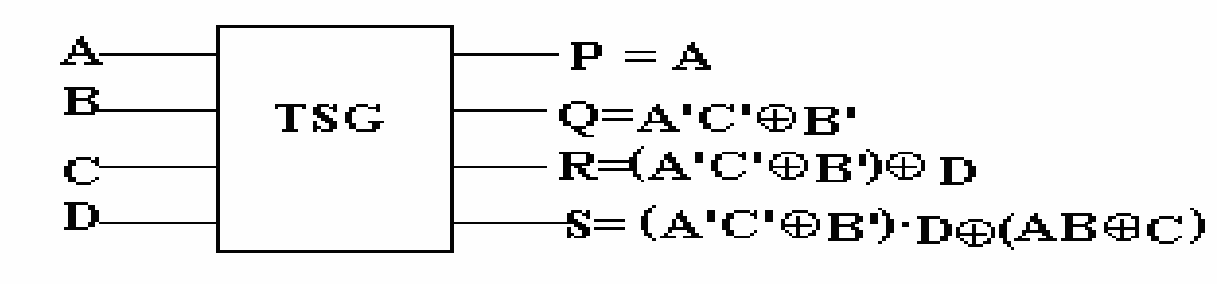

Figure 1. Reversible TSG Gate

One of the important functions of the TSG gate is that it can work as a reversible full adder unit. Fig. 2 shows the TSG gate configured thus. A full adder designed using a TSG gate is the most optimal in terms of the numbers of reversible gates and 'garbage outputs'. (A garbage output is an output that is not used in further computations.)

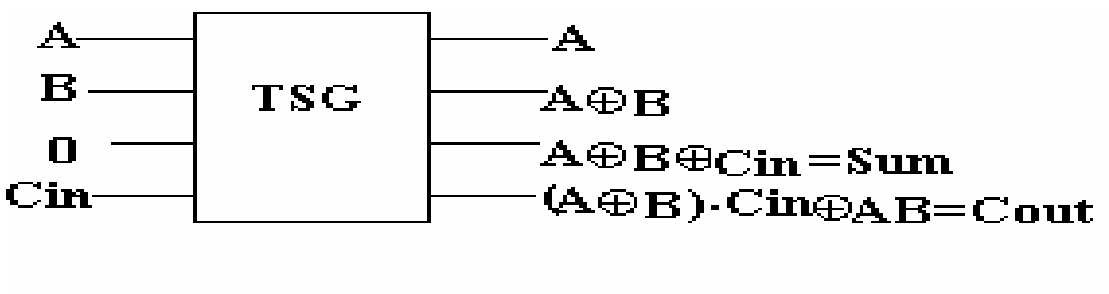

Figure 2. Reversible TSG Gate as Full Adder

A number of reversible full adders were proposed in [15,16,17,18]. The reversible full adder circuit in [15] requires three reversible gates (two 3*3 New gates and one 2*2 Feynman gate) and produces three garbage outputs. The reversible full adder circuit in [16,17] requires three reversible gates (one 3*3 New gate, one 3*3 Toffoli gate and one 2*2 Feynman gate) and produces two garbage outputs. The design in [18] requires five reversible Fredkin gates and produces five garbage outputs. The full adder designed using TSG in Fig. 2 requires only one reversible gate (one TSG gate) and produces only two garbage outputs. Hence, the full-adder design in Fig. 2 using TSG gate is better than the previous full-adder designs of [15,16,17,18]. A comparison is shown in Table I.

TABLE I. COMPARISON OF REVERSIBLE FULL ADDERS

| | No of Gates | No of Garbage Outputs | Unit Delay |
|---|---|---|---|
| Full adder Using TSG | 1 | 2 | 1 |
| Existing Circuit[15] | 3 | 3 | 3 |
| Existing Circuit [16,17] | 3 | 2 | 3 |
| Existing Circuit[18] | 5 | 5 | 5 |

## III. REVERSIBLE LOGIC IN HARDWARE CRYPTOGRAPHY

The ALU of a crypto-processor is the major source of power consumption. It consists of carry propagate adders; carry save adders, multipliers, squares, registers and accumulators. Thus, the ALU of a crypto-processor has been designed using reversible logic. As a prototype of applying reversible logic for designing the crypto-ALU, reversible designs of carry propagate adders and five-to-two CSA and four-to-two CSA adders [6] are presented. Figure 3 shows the reversible carry propagate adder using the TSG. In a public key encryption such as RSA, 1024 bit addition is performed and this requires carry propagate as well as carry save adders. Recently, for conventional hardware cryptography, the authors have demonstrated one of the fastest designs of Montgomery multiplication using compressors and carry save adders [19]. The proposed reversible designs in this paper are also applicable to design the fastest reversible equivalent of Montgomery multiplication.

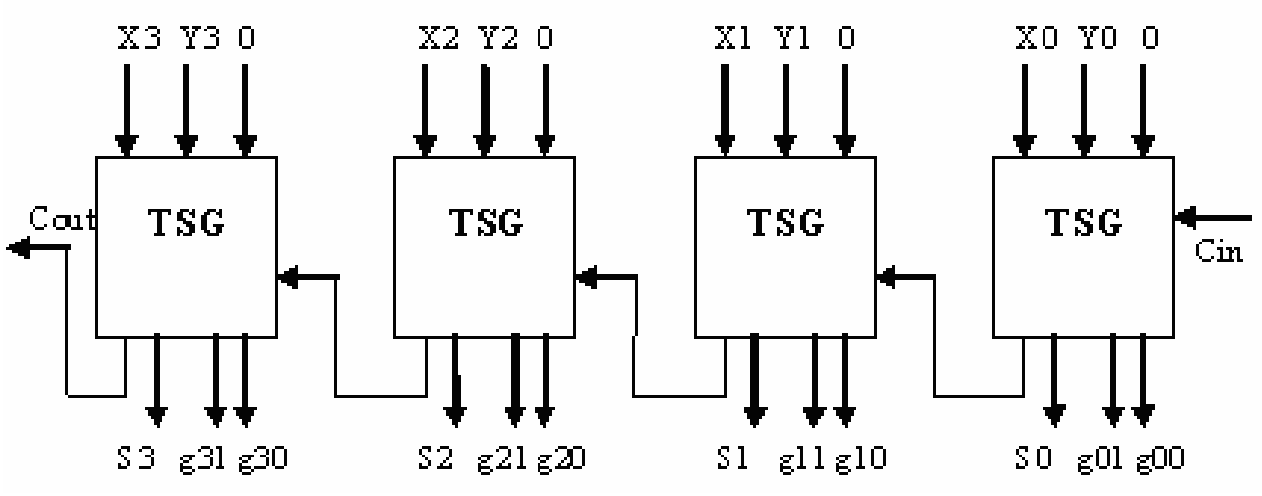

Figure 3. Reversible Carry Propagate Adder

Figure 4 and Figure 5 show reversible designs for four-to-two and five-to-two CSAs. The reversible CSA circuits are needed for zero power dissipating Montgomery [9] modulo multipliers.

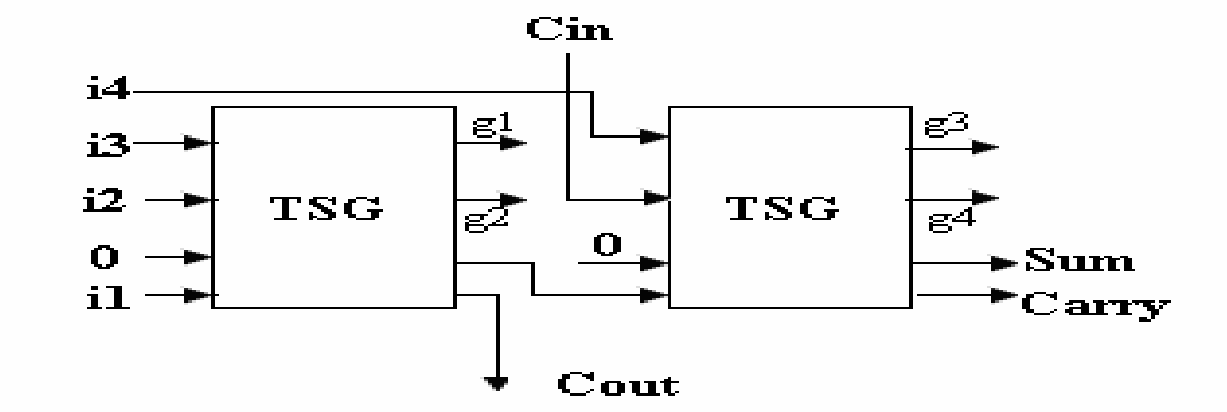

Figure 4. Four-to-two CSA Using Reversible Logic

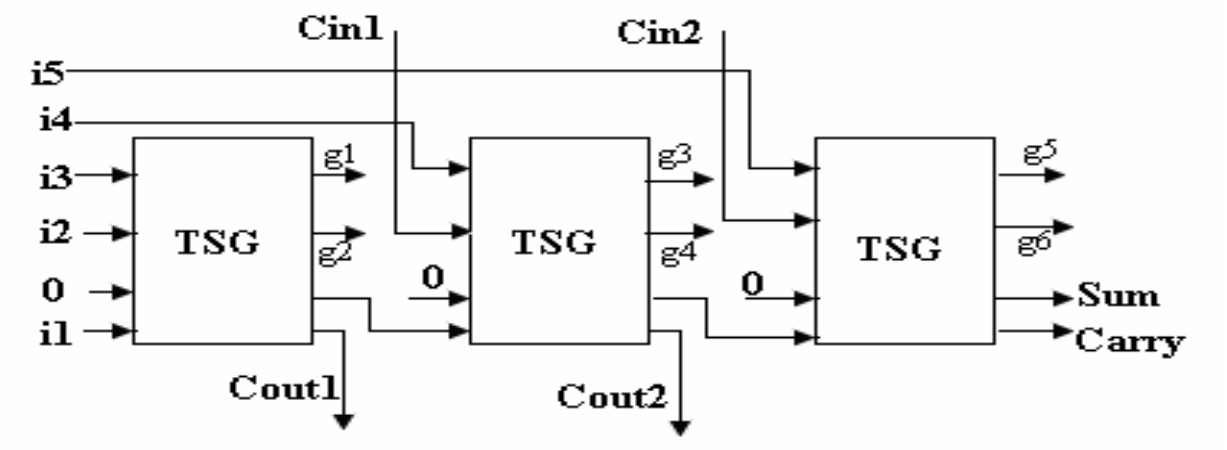

Figure 5. Five-to-two CSA Using Reversible Logic

The Montgomery multiplication algorithm is used for modulo multiplication in hardware cryptosystems. The modified Montgomery multiplication algorithm [32] is shown below.

Algorithm 1: Montgomery Multiplication
Inputs: $X,Y,M$ with $0 \leq X,Y < M$
Output: $P = (X \times Y \times (2n)\text{-}1) \bmod M$
$n$: number of bits in $X$,
$xi$ : $ith$ bit of $X$
$s0$: LSB of $S$

1. $S := 0$; $C := 0$;
2. **for** $i$:=0 **to** $k$-1 **do**
3. $S,C := S + C + xi*Y$;
4. $S,C := S + C + s0*M$;
5. $S := S$ div 2; $C := C$ div 2;
6. $P := S + C$
7. **if** $P \geq M$ **then** $P := P - M$;

It is evident that in addition to CSA adders, reversible sequential components like registers and shift registers will also be required for the reversible Montgomery multiplier. Thus, the design of reversible sequential circuits is also proposed in this paper as shown below.

### *A. Proposed Reversible Sequential Circuits*

Firstly, the reversible D latch is built from the Fredkin gate which is later used to design complex sequential circuits, as discussed in the section below. We have previously also proposed designs of reversible sequential circuits. As far as we can discover, this is the first work to design reversible sequential circuits [20,21]. The proposed reversible sequential designs are further modifications to the existing design previously proposed. The Fredkin gate [22], is a (3*3) conservative reversible gate originally introduced by Petri. It is called a 3*3 gate because it has three inputs and three outputs. The term *conservative* means that the Hamming weight (number of logical ones) of its input equals the Hamming weight of its output. The input triple $(x_1, x_2, x_3)$ generates the output triple $(y_1, y_2, y_3)$ as follows:

$$y_1 = x_1$$

$$y_2 = \bar{x}_1 \cdot x_2 + x_1 \cdot x_3$$

$$y_3 = x_1 \cdot x_2 + \bar{x}_1 \cdot x_3$$

Because the circuit is reversible, *x* and *y* can be interchanged.

### *B. Proposed Reversible D Latch Using Fredkin Gate*

Figure 6 shows a conventional D latch. The characteristic equation of the D latch can be written as

$$Q^+ = D \cdot E + \bar{E} \cdot Q .$$

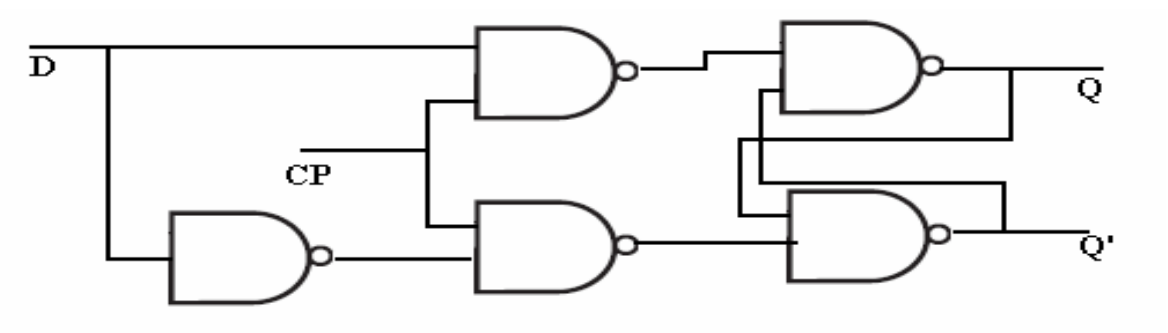

Figure 6. Conventional D Latch

The characteristic equation of the D latch can be mapped onto the Fredkin gate (F). Figure 7 shows the realization of the proposed reversible D latch. To avoid a fan-out problem, a Feynman gate (FG) [22] is used to copy the output. It can be seen that the reversible D latch is highly optimized in terms of the number of reversible gates and garbage outputs (outputs that are not subsequently used).

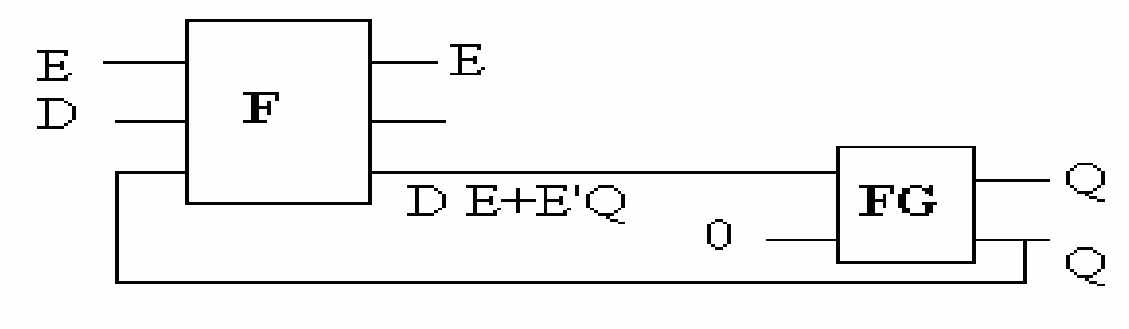

Figure 7. Reversible D latch built from a Fredkin Gate

### *C. Complex Sequential Circuits*

The reversible D latch is used to implement more complex reversible sequential circuits. Figure 8 shows a reversible storage register constructed from four reversible D latches and a common clock input. Figure 9 shows the master-slave D flip flop designed from the proposed D Latch, in which CP refers to the clock pulse.

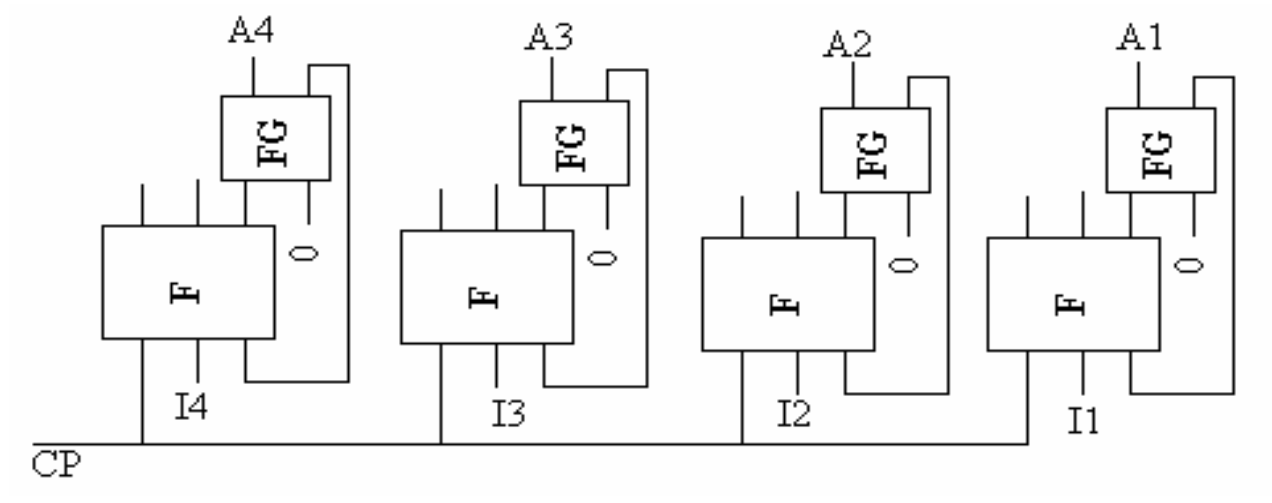

Figure 8. Reversible Register built from reversible D latches

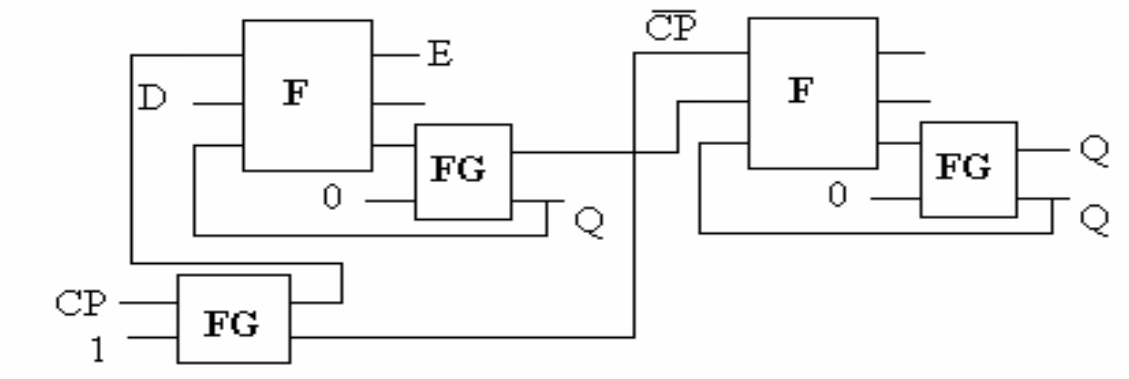

Figure 9. Reversible Master Slave D Flip Flop

Figure 10 shows a reversible shift register built from the reversible master-slave D flip flop. Each clock pulse shifts the contents of the register one bit position to the right.

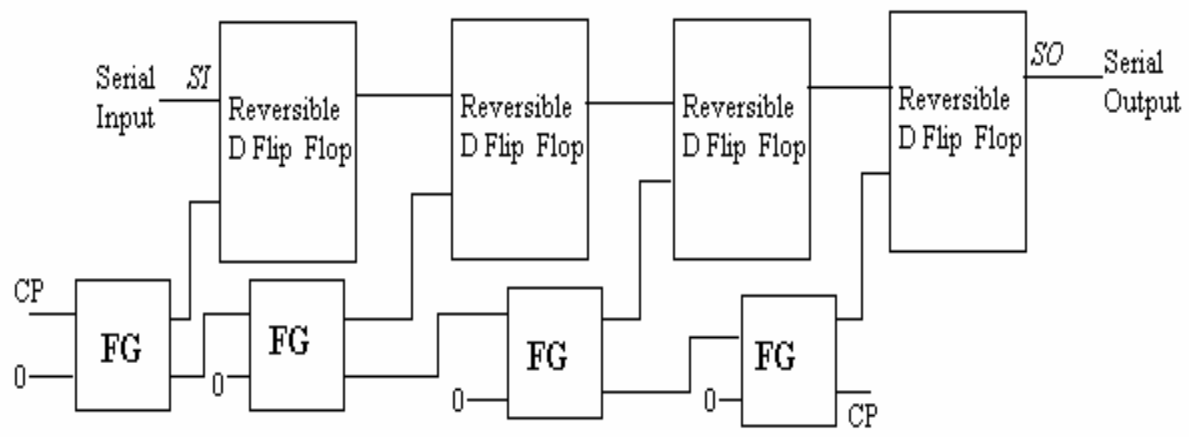

Figure 10. Reversible Shift Register built from reversible master-slave D flip flop

### *D. Reversible Montgomery Multiplier*

Figure 11 shows the reversible implementation of the Montgomery multiplier using the reversible components shown in this paper. The first reversible CSA in Figure 11 performs S,C := S + C + xi*Y; the S,C produced are stored in proposed reversible registers S and C. The LSB $S_0$ of the reversible register S is multiplied with M to generate $S_0$ * M. After its generation, the second CSA performs S,C := S + C + $s_0$*M. The S,C thus generated are passed to the reversible shift registers to produce S div 2 and C div 2. The other components required in the multiplier such as for (P := S + C and if P ≥ M then P := P – M) can also be designed using reversible logic.

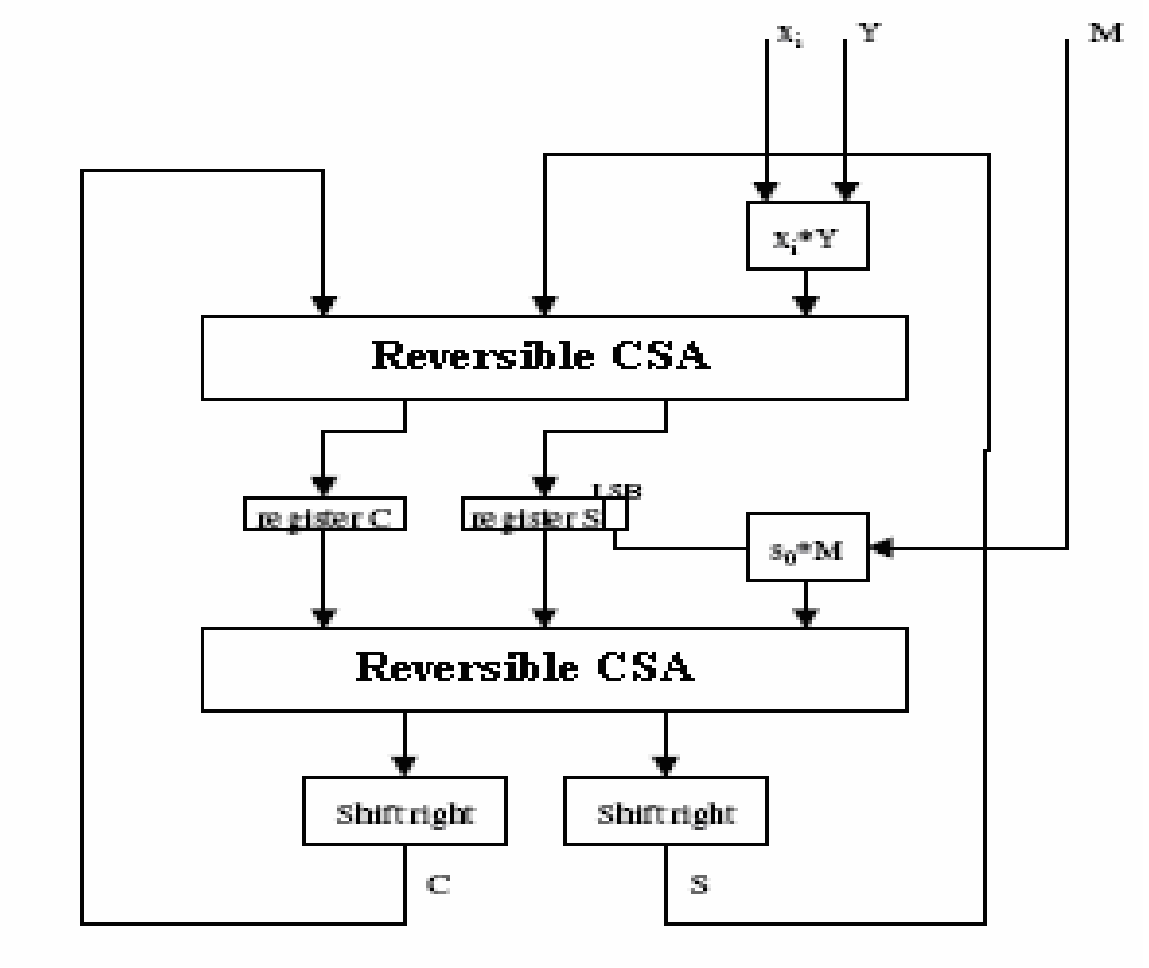

Figure 11. Reversible circuitry of Montgomery multiplication ($x_i$ * Y and $S_0$ * M are reversible AND operations)

Another possible way to implement Montgomery multiplication, which is assumed to be the best choice for hardware on current integrated circuits (ICs) is systolic array. An impressive work to design Montgomery multiplier using systolic array is proposed in [23]. In [23], it is demonstrated that in the systolic array to design Montgomery multiplier, there are basically four different types of cells. These cells basically consist of full adders and half adders. Thus, in order to demonstrate the design of systolic cells using reversible logic, the authors are proposing their optimal reversible designs best in terms of number of reversible gates and garbage outputs. It is to be noted that TSG gate is the best gate to design reversible full adder but if we use TSG gate to design reversible half adder, the garbage outputs will increase in the proposed design. Thus, we have used New Gate [15] to design reversible half adder to make the design highly optimized in terms of number of reversible gates and garbage outputs. The New gate can realize the half adder with bare minimum of one garbage output. Figure 12 (a) shows the New Gate and Figure 12(b) shows its working as a reversible half adder.

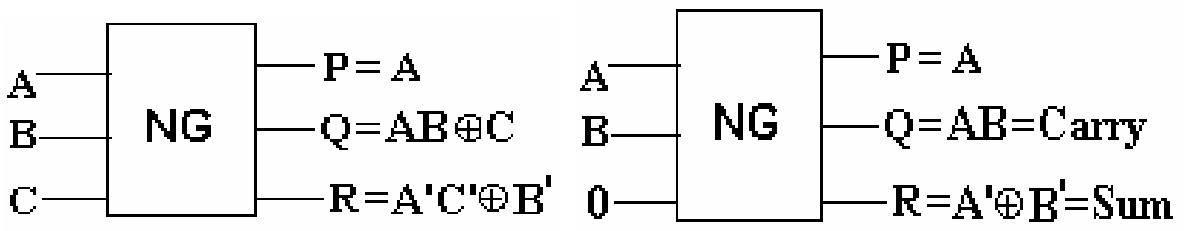

(a) (b)

Figure 12. (a) New Gate(NG) (b) New Gate as Half Adder

The proposed reversible systolic cells are shown in Figure 13(NG is New Gate, F is Fredkin Gate and FG is Feynman Gate). The optimality of the proposed designs can be easily understood from Table II which shows the comparative study using different existing full adder circuits proposed in literature. The proposed optimal reversible systolic cells can later be used to design efficient and optimal reversible systolic array and Montgomery multiplier as demonstrated in [23]. Due to limitation of space, the details of systolic cells are not discussed, which can be found in [23] for irreversible counterparts. The main idea is to bring the attention of crypto experts towards an alternative way of DPA resistant cryptographic hardware and how to design them in an optimal manner.

TABLE II. COMPARISON OF REVERSIBLE REGULAR SYSTOLIC CELLS

| | No of Gates | No of Garbage Outputs | Unit Delay |
|---|---|---|---|
| Full adder Using TSG | 5 | 9 | 4 |
| Existing Circuit[15] | 9 | 11 | 8 |
| Existing Circuit [16,17] | 7 | 9 | 7 |
| Existing Circuit[18] | 13 | 15 | 13 |

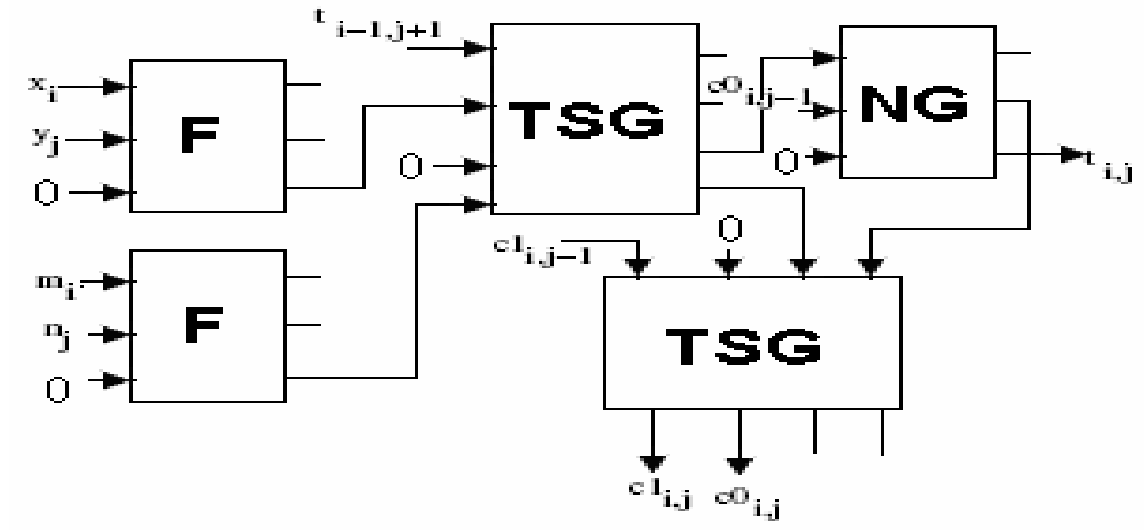

**(a)**

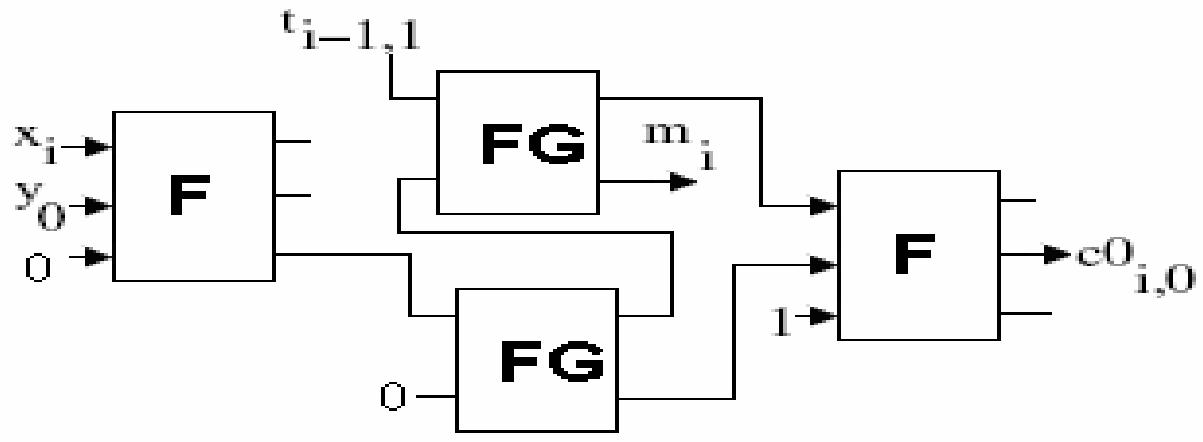

**(b)**

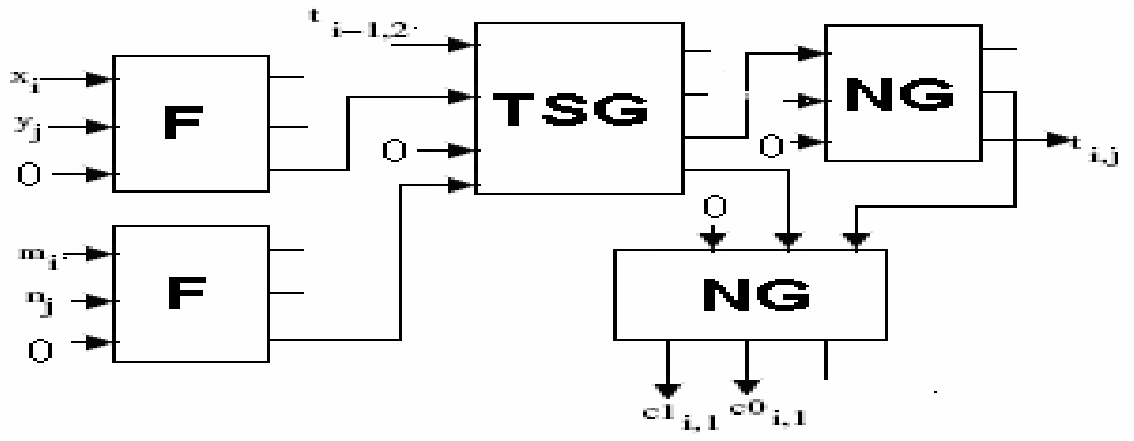

**(c)**

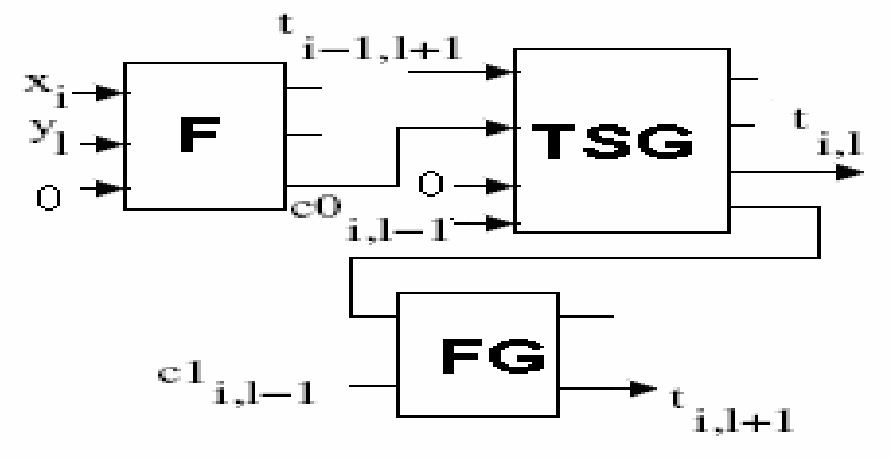

**(d)**

Figure 13. Reversible Designs of systolic array cells; (a) Regular, (b) Rightmost, (c) 1-st bit, (d) Leftmost

### *E. Importance of Reversible Montgomery Multiplication*

The importance of low power Montgomery multiplication can be understood from Algorithm 2 which demonstrates the algorithm employed to compute $a^b$ mod n(RSA encryption and decryption functions ).

Algorithm 2: *Algorithm to compute $a^b$ mod n efficiently*

```
l = 0; m =1
  For j = k downto 0
      do l = 2*l
          m = MontMult (m, m, n)  // Montgomery Multiplier
          if bj = 1
               then   l= l+1
             m =MontMult (m, a, n) // Montgomery Multiplier
 return m
```

From algorithm 2, it is evident that the Montgomery multiplier is called k+1 times in the loop of RSA encryption and decryption functions. Thus, a low power reversible Montgomery Multiplier is the immediate requirement of RSA and similarly ECC encryption systems to design DPA resistance encryption chips. It is to be noted that in the design of the cryptographic hardware that is resistant to Differential Power Analysis (DPA), there is no need to make the cryptographic hardware completely reversible. Only those parts which are the main source of power consumption have to be made reversible, other components such as control logic can also be designed with conventional irreversible logic.

## IV. CONCLUSION

This paper proposes the novel idea of applying reversible logic for the design of secure cryptosystems. The reversible design of carry propagate adder, five-to-two CSA and four-to-two CSA adders has been demonstrated. Novel reversible sequential circuits are also proposed. The proposed adder and sequential units are used to design an efficient modular Montgomery multiplier for use in hardware cryptosystems. The future work in this direction is the comprehensive implementation of DES or RSA using reversible logic, and to provide a rigorous analysis. It is suggested that the proposed work will provide a new focus in the cryptography field making hardware more secure against DPA attacks and will also attract the attentions of computer scientists towards applying reversible logic to hardware cryptography.